\documentstyle[aps,preprint]{revtex}
\begin{document}

\draft


\preprint{YITP-99-30}

\title{Quasi-local first law of black-hole dynamics}
\author{Shinji Mukohyama${}^{\dagger}$
and
Sean A. Hayward${}^{\ddagger}$}
\address{
${}^{\dagger,\ddagger}$Yukawa Institute for Theoretical Physics, 
Kyoto University \\
Kyoto 606-8502, Japan \\
${}^{\dagger}$Department of Physics and Astronomy, 
University of Victoria \\
Victoria, BC, V8W 3P6, Canada\\
${}^{\ddagger}$
Center for Gravitational Physics and Geometry, 
104 Davey Laboratory\\ 
The Pennsylvania State University\\
University Park, PA16802-6300, USA
}
\date{\today}

\maketitle


\begin{abstract} 

A property well known as the first law of black hole is a relation
among infinitesimal variations of parameters of stationary black
holes. We consider a dynamical version of the first law, which may be
called the first law of black hole dynamics. The first law of black
hole dynamics is derived without assuming any symmetry or any
asymptotic conditions. In the derivation, a definition of dynamical
surface gravity is proposed. In spherical symmetry it reduces to that
defined recently by one of the authors (SAH).

\end{abstract}



Black hole thermodynamics, analogies between the theory of black holes
and thermodynamics, has been one of the hottest fields in black hole 
physics since Bekenstein's introduction of black hole
entropy~\cite{Bekenstein}. The black hole entropy was introduced as a
quantity proportional to horizon area. The proportionality coefficient
was fixed by Hawking's discovery that a black hole with surface
gravity $\kappa$ emits radiation with temperature
$T_H=\kappa/2\pi$~\cite{Hawking1975}: by identifying $T_H$ with the 
temperature of the black hole, black hole entropy is determined to be
one quarter of the horizon area. The expression of black hole entropy
is called the Bekenstein-Hawking formula.

To determine the coefficient in the Bekenstein-Hawking formula
from the expression of the Hawking temperature $T_{H}$, the first law
of black holes~\cite{BCH} is used. The first law is a relation among
infinitesimal variations of parameters of stationary black holes:
horizon area, mass, angular momentum, etc. 
Strictly speaking, it does not relate dynamical evolutions of these
quantities. Thus, it might be physically non-trivial to connect the
temperature of dynamically emitted radiation with black hole entropy
by using the first law. Nonetheless, the Bekenstein-Hawking formula
obtained by using the first law is acceptable. For example, in
Euclidean gravity the Bekenstein-Hawking formula for a Schwarzschild
black hole is correctly obtained by requiring a regularity of the
corresponding Euclidean section~\cite{Gibbons&Hawking}.

Moreover, the first law is used in (quasi-stationary but) dynamical
situations to prove the generalized second
law~\cite{Frolov&Page,Mukohyama-GSL}, which is a natural 
generalization of both the second law (or area law~\cite{Hawking1971}) of
black holes and  the second law of usual thermodynamics. 
In the proof, by assuming quasi-stationarity, the use of the first law
is extended to relate small changes of physical quantities from an
initial near-stationary black hole to a final near-stationary
one. However, this idea of quasi-stationarity is an approximation; if
we intend to prove the generalized second law for finite changes
between initial and final near-stationary black holes or to a purely
dynamical situation, the stationary first law can not be used.

Therefore, to make black hole thermodynamics self-consistent it
must be shown that a dynamical version of the first law of black holes 
exists. In Ref.~\cite{U1Law}, it was derived assuming spherical
symmetry and may be called the first law of black hole dynamics. The
purpose of this paper is to derive the first law of black hole
dynamics {\it without assuming any symmetry or any asymptotic
conditions}.


In this paper we treat a dynamical and not necessarily asymptotically 
flat spacetime. Even for such a general situation, there is a definition 
of a black hole as a certain type of trapping
horizon~\cite{BH-dynamics}. A {\it trapping horizon} is a
three-surface foliated by marginal surfaces, where a {\it marginal
surface} is a spatial two-surface on which one null normal expansion
defined below vanishes. 
Geometrically this is where a light wave would have instantaneously
parallel rays. The physical idea is that gravity can trap an expanding 
light wave and make it contract. 
We mention here that different types of trapping horizon can be
regarded as defining a black hole, a white hole or a
wormhole~\cite{Wormhole}. However, the distinctions are irrelevant for
the purpose of this paper. The first law we shall obtain holds for any
trapping horizon.


To investigate the behavior of the trapping horizon, the so-called 
double-null formalism or $(2+2)$ decomposition of general
relativity is useful. Among several
$(2+2)$-formalisms~\cite{Double-null,Hayward(2+2)}, we adopt one 
based on Lie derivatives w.r.t null vectors 
developed by one of the authors~\cite{Hayward(2+2)}. 
Let us review basic ingredients
of the formalism. Suppose that a four-dimensional spacetime manifold 
$(M,g)$ is foliated (at least locally) by null
hypersurfaces $\Sigma^{\pm}$, each of which is parameterized
by a scalar $\xi^{\pm}$, respectively. 
The null character is described by $g^{-1}(n^{\pm},n^{\pm})=0$, 
where $n^{\pm}=-d\xi^{\pm}$ are normal $1$-forms to $\Sigma^{\pm}$.
The relative normalization of the null normals defines 
a function $f$ as $g^{-1}(n^+,n^-)=-e^f$.
The intersections of $\Sigma^+(\xi^+)$ and
$\Sigma^-(\xi^-)$ define a two-parameter family of two-dimensional
spacelike surfaces $S(\xi^+,\xi^-)$. Hence, by introducing an intrinsic
coordinate system ($\theta^1$,$\theta^2$) of the $2$-surfaces, 
the foliation is described by the imbedding 
$x = x(\xi^+,\xi^-;\theta^1,\theta^2)$.

For the imbedding, the intrinsic metric on the $2$-surfaces is found 
to be $h=g+e^{-f}(n^+\otimes n^- +n^-\otimes n^+)$. 
Correspondingly, the vectors $u_{\pm}=\partial/\partial\xi^{\pm}$ 
have 'shift vectors' $s_{\pm}=\perp u_{\pm}$, where $\perp$ indicates 
projection by $h$.
The $4$-dimensional metric is written in terms of ($h$,$f$,$s_{\pm}$) 
as
%
\begin{equation}
 g = \left(
	\begin{array}{ccc}
	h(s_+,s_+) &	h(s_+,s_-)-e^{-f} &	h(s_+) \\
	h(s_-,s_+)-e^{-f} &	h(s_-,s_-) &	h(s_-) \\
	h(s_+) &	h(s_-) &	h 
	\end{array}
	\right).
\end{equation}
Geometrical quantities such as {\it expansions} $\theta_{\pm}$, 
{\it shears} $\sigma_{\pm}$ and 
the {\it twist} $\omega$ are defined by 
$\theta_{\pm}=*{\cal L}_{\pm}*1$, 
$\sigma_{\pm}=\perp{\cal L}_{\pm}h-\theta_{\pm}h$ and 
$\omega=e^f [l_-,l_+]/2$, where
$*$ denotes the Hodge-dual operator of $h$,
$l_{\pm}=u_{\pm}-s_{\pm}=e^{-f}g^{-1}(n^{\mp})$ 
are null normal vectors to $\Sigma^{\pm}$, 
and ${\cal L}_{\pm}$ denotes the Lie derivative along $l_{\pm}$, 
respectively. 
It is possible to write down the Einstein tensor in terms of these
geometrical quantities. The component useful for our purpose is
$G_{+-}=G(l_+,l_-)$, which is given by
%
\begin{equation}
 2e^fG_{+-} = 
	{}^{(2)}R + 
	e^f ({\cal L}_+\theta_- +{\cal L}_-\theta_++2\theta_+\theta_-)
	-2\left[ h(\omega,\omega) + \frac{1}{4}h^{\sharp}(df,df)
	\right] + {\cal D}^2f.
\label{eqn:G}
\end{equation}
Here $h^{\sharp}=g^{-1}hg^{-1}$ is $h$ raised by $g^{-1}$, 
${\cal D}^2$ and ${}^{(2)}R$ are the two-dimensional Laplacian  
and the Ricci scalar associated with the metric $h$.


Before deriving the first law we have to define energy and surface
gravity in a quasi-local way. In spherical symmetry there 
is a widely accepted energy: the Misner-Sharp (MS) 
energy~\cite{Misner-Sharp}. In Ref.~\cite{U1Law} the MS energy is 
used to derive the first law of black hole dynamics in spherical symmetry. In this 
paper we adopt the Hawking energy~\cite{Hawking-energy}, which
reduces to the MS energy in spherical symmetry. It is defined by
%
\begin{equation}
 E(\xi^+,\xi^-) = \frac{r}{16\pi}\int_{S(\xi^+,\xi^-)} 
        d^2\theta\sqrt{h} 
        \left[{}^{(2)}R + e^{f}\theta_+\theta_-\right],
        \label{eqn:energy}
\end{equation}
where $h$ is the determinant of the two-dimensional metric $h_{ab}$
and the area radius $r$ is defined by 
%
\begin{equation}
 r = \sqrt{A/4\pi},
 A = \int_{S(\xi^+,\xi^-)} d^2\theta\sqrt{h}.
\end{equation}


In Ref.~\cite{U1Law}, a definition of dynamical surface gravity was 
proposed in spherical symmetry. A natural generalization to a not 
necessarily spherically symmetric case is 
%
\begin{equation}
 \kappa (\xi^+,\xi^-) = \frac{-1}{16\pi r}
	\int_{S(\xi^+,\xi^-)}d^2\theta\sqrt{h} \ 
        e^{f}
	({\cal L}_+\theta_-+{\cal L}_-\theta_++\theta_+\theta_-). 
	\label{eqn:kappa}
\end{equation}
This is the most simple generalization in the sense that 
it includes neither the shear $\sigma_{Aab}$ nor 
the twist $\omega^a$.


We now derive the first law of black hole dynamics for the Hawking energy 
(\ref{eqn:energy}) and the surface gravity defined by 
(\ref{eqn:kappa}). 
It is easy to show that 
%
\begin{equation}
 dE - \frac{\kappa}{8\pi} dA = 
        wAdr + r d\left(\frac{E}{r}\right),\label{eqn:pre-1st-law}
\end{equation}
where $w$ is defined by
%
\begin{equation}
 w = \frac{1}{A}
        \left(\frac{E}{r} 
        -\kappa r\right).
\end{equation}
Here note that '$d$' in Eq.~(\ref{eqn:pre-1st-law}) is not a variation 
in a space of stationary solutions of the Einstein equation as in the
first law of black hole statistics, but is the differentiation w.r.t.
the parameters $\xi^{\pm}$ of the spacetime foliation. 
(For example, $dE=d\xi^+\partial_+E+d\xi^-\partial_-E$.) 
We mention that Eq.~(\ref{eqn:pre-1st-law}) holds independently of the 
definitions of $E$ and $\kappa$ while the following arguments depend 
on the definitions.

A marginal surface is a surface where one of the expansions
$\theta_{\pm}$ vanishes. Since the Gauss-Bonnet theorem says that 
%
\begin{equation}
 \int_{S(\xi^+,\xi^-)} d^2\theta\sqrt{h} \ {}^{(2)}R = 
        8\pi (1-\gamma ),
\end{equation}
where $\gamma$ is the genus or number of handles of $S(\xi^+,\xi^-)$, 
the energy divided by area radius is given by $E/r=(1-\gamma )/2$ on a 
marginal surface and is a constant. Thus, 
%
\begin{equation}
 E' = \frac{\kappa}{8\pi} A' + wAr',
\end{equation}
where the prime denotes the derivative along the trapping horizon. 
This is the first law of black hole dynamics. Note that this also
holds along any hypersurface foliated by 2-surfaces on which $E/r$ is
constant.


By using Eq.~(\ref{eqn:G}) it is easy to show that $w$ is written as 
follows.
%
\begin{equation}
 w = w_m + w_j, 
\end{equation}
where the averaged matter energy density $w_m$ and the effective
angular energy density $w_j$ are defined by 
%
\begin{eqnarray}
 w_m & = & \frac{1}{8\pi A}
        \int_{S(\xi^+,\xi^-)}d^2\theta\sqrt{h}  
        e^{f}G_{+-},\nonumber\\
 w_j & = & \frac{1}{8\pi A}
        \int_{S(\xi^+,\xi^-)}d^2\theta\sqrt{h}\left[
        h(\omega,\omega) + \frac{1}{4}h^{\sharp}(df,df) \right].
\end{eqnarray}
The Einstein equation $G=8\pi T$ says that $w_m$ is $e^fT(l_+,l_-)$
averaged over the $2$-surface. It seems that $w_j$ represents
effective energy density due to deviation from spherical symmetry
(eg. angular momentum). 

The term $wAr'$ should be a work term done along the horizon. For 
example, for an electromagnetic field, the term $w_mAr'$ reduces to 
the electromagnetic work done along the horizon~\cite{U1Law}. It seems 
that the term $w_jAr'$ is a work associated with deviation from
spherical symmetry (eg. angular momentum) of the trapping horizon.


In this paper the first law of black hole dynamics has been derived 
without assuming any symmetry or any asymptotic condition. 
In the derivation we have given a new 
definition of dynamical surface gravity. In spherical symmetry it 
reduces to that defined in Ref.~\cite{U1Law}.

Now some comments are in order. 
First, besides the first law derived in this paper, there exist the 
second law~\cite{BH-dynamics} and perhaps a third law~\cite{3rd-law}
for the trapping horizon (or apparent horizon). It seems that by using 
these laws we can formulate black hole thermodynamics consistently as 
trapping horizon dynamics. However, for this purpose, 
there is an important open question:
we have to associate temperature of quantum fields with the 
trapping horizon. 
All we can say here is that the temperature may be given by 
$\hbar\kappa/2\pi$, where $\kappa$ is the surface gravity introduced in 
this paper.

The second and the third (final) comments are on the definition 
(\ref{eqn:kappa}) of the surface gravity. 
By definition, $r\kappa$ is not a 
local quantity but a quasi-local quantity defined as an integral 
over the surface $S(\xi^+,\xi^-)$. However, it is expected that the 
integrand is constant over the integration surface 
under certain conditions of equilibrium. Although such a statement has 
not been established, it may be called the zeroth law of black hole 
dynamics if it is proved in some situation.

A final comment is in order. The surface gravity $\kappa(\xi^+,\xi^-)$ 
is an invariant of a double-null foliation at the surface. 
Since a non-null trapping horizon locally determines a unique 
double-null foliation, the surface gravity is an 
invariant of the trapping horizon if the horizon is not null.
On the other hand, the null case is ambiguous because of the freedom to 
rescale the other null direction.  
Fixing this would require some kind of limiting argument that might be 
effectively a zeroth law.
Therefore, we have to impose an auxiliary condition 
for the surface gravity $\kappa(\xi^+,\xi^-)$ to work well 
when the trapping horizon is null.
Since surface gravity seems to be related to temperature of quantum 
fields as stated above, it will be valuable to investigate the auxiliary 
condition in detail.

\vskip 1cm

\centerline{\bf Acknowledgments}

We would like to thank Professor H. Kodama and Professor W. Israel for
their continuing encouragement. S.M. was supported by JSPS Research
Fellowships for Young Scientists, and this work was supported
partially by the Grant-in-Aid for Scientific Research Fund
(No. 9809228).


\end{document}